\documentclass[onecolumn,10pt]{article}
\usepackage{amsmath}
\usepackage{amsfonts}
\usepackage{amssymb}
\usepackage{amsthm}
\usepackage{times}
\usepackage{cancel}
\usepackage{graphics}

\newcommand{\hil}[1]{\mbox{$\mathcal{#1}$}}

\newcommand{\ket}[1]{| #1 \rangle}

\usepackage{soul}
\usepackage{tikz}

\title{\textbf{Bananaworld: \\ Quantum Mechanics for Primates}}
\author{Jeffrey Bub\\ \small \textit{Philosophy Department and Institute for Physical Science and Technology}\\  \small \textit{University of Maryland, College Park, MD 20742, USA}}
\date{}
\normalsize
\begin{document}

\maketitle

\bigskip
\bigskip

\begin{abstract}
This is intended to be a serious paper, in spite  of the title. The idea is that quantum mechanics is about probabilistic correlations, i.e., \emph{about the structure of information}, since a theory of information is essentially a theory of probabilistic correlations. To make this clear,  it suffices to consider measurements of  two binary-valued observables, $x$ with outcomes $a=0$ or 1, performed by Alice in a region $\mathbf{A}$, and $y$ with outcomes $b=0$ or 1 performed by Bob in a separated region $\mathbf{B}$---or, to emphasize the banality of the phenomena, two ways of peeling a banana, resulting in one of two tastes. The imagined bananas of Bananaworld are non-standard, with operational or phenomenal probabilistic correlations for peelings and tastes that lie outside the polytope of local correlations. The `no go' theorems tell us   that we can't shoe-horn these correlations into a classical correlation polytope, which has the structure of a simplex, by supposing that something has been left out of the story, without giving up fundamental principles that define what we mean by a physical system. The nonclassical features of quantum mechanics, including the irreducible information loss on measurement, are shown to be generic features of correlations that lie outside the local correlation polytope. As far as the conceptual problems are concerned, we might as well talk about bananas.
\end{abstract}

\bigskip
\bigskip

This paper is intended to be serious, in spite of the title. The idea is that quantum mechanics is about probabilistic correlations, i.e., \emph{about the structure of information}, since a theory of information is essentially a theory of probabilistic correlations---not about energy being quantized in discrete lumps or quanta, not about particles being wavelike, not about the universe continually splitting into countless co-existing quasi-classical universes, with many copies of ourselves, or anything like that. To make this clear,  it suffices to consider measurements of  two binary-valued observables, $x$ with outcomes $a=0$ or 1, performed by Alice in a region $\mathbf{A}$, and $y$ with outcomes $b=0$ or 1 performed by Bob in a separate region $\mathbf{B}$. In fact, it suffices to consider bananas.

In the first section (`Bananaworld'), I show that a variety of `no go' theorems about the impossibility of extending quantum mechanics---by Bell \cite{BellEPR}, Kochen-Specker \cite{KochenSpecker}, Pusey-Barrett-Rudolph \cite{PBR2012}, Colbeck-Renner \cite{ColbeckRenner2012}---can be applied to operational or phenomenal correlation phenomena involving bananas on an imaginary island called Bananaworld. The point is to make clear that conceptual puzzles associated with the completeness problem---the `hidden variable' problem, broadly construed---are features of rather banal correlation phenomena that have nothing in particular to do with Hilbert space or the specific behavior of microsystems evolving according to the wave dynamics of the Schr\"{o}dinger equation. 

The second section (`Taking Stock') develops the notion of probabilistic correlations in terms of correlation arrays. Operational correlations represented by points inside the polytope of local correlations can be simulated with classical resources, which generate classical correlations represented by the points in a simplex, where the vertices of the simplex represent joint deterministic states that are the common causes of the correlations. The conceptual problems in Bananaworld, as in our quantum world, arise because probabilistic correlations can lie outside the local correlation polytope. In Bananaworld, as in other non-classical (= non-simplex) worlds, cloning an arbitrary extremal state is impossible without violating the no-signaling principle, and there is a necessary information loss associated with measurement.  

In the third section (`The Measurement Problem'), I consider the measurement problem in Bananaworld. There are two distinct aspects to the measurement problem. In a simplex theory, it is possible to  explain dynamically how a particular measurement outcome is selected in a measurement process. In a non-simplex theory, any such dynamical account would be inconsistent with the  no-signaling principle. The selection of a particular outcome in a quantum measurement, and the associated loss of complementary information, is a genuinely random event, a `free choice' on the part of the system, not the culmination of a dynamical process. So there is nothing further to be said about this aspect of measurement in a non-simplex theory. The second aspect is the problem of how to account for the definiteness of the part of the universe that records the outcomes of quantum measurements or random Bananaworld events. We can't simply assume that quantum phenomena or Bananaworld phenomena take place in a classical arena that provides the physical infrastructure for definite stable records. I argue that this is a consistency problem, and I propose a possible solution. 

\section{Bananaworld}
Imagine discovering an island with banana trees---Bananaworld. The bananas in Bananaworld are peculiar in that half the peel is the familiar yellow, and the other half brown, with a seam connecting the two halves. There are two ways to peel one of these bananas: one can either peel the yellow half ($Y$) along the seam, or the brown half ($B$). A Bananaworld banana tastes just like an \emph{o}rdinary banana (0), or the flavor is \emph{i}ntense, \emph{i}ncredible, \emph{i}ndescribably delicious  (1), seemingly dependent on how you peel the banana. 

So there are two things you can do to a banana in Bananaworld (corresponding to a binary choice of measurements), and two possible outcomes (corresponding to two possible measurement outcomes).

It turns out that there are various sorts or states of bananas. Some trees have a large bunch of bananas that one might call `pure state' bananas. There are four types of pure state bananas: $Y0, Y1, B0, B1$. For all the bananas on a $Y0$ bunch, if you peel the yellow half, the banana tastes ordinary, but if you peel the brown half, the banana tastes ordinary  or intense with probability 1/2. Similarly, for all the bananas on a $Y1$ bunch, if you peel $Y$, the banana tastes intense , but if you peel $B$, the banana tastes ordinary  or intense  with probability 1/2. And similarly for $B0$ and $B1$ bananas.

There are also trees that have one bunch with just two bananas. If you peel $Y$ or $B$ for either of these bananas, the banana tastes ordinary (0) or intense (1) with equal probability. But peelings and tastes for such a banana pair are correlated like the correlations of a Popescu-Rohrlich box (PR-box) \cite{PopescuRohrlich94}, and the correlations persist if the bananas are separated and then peeled when they are apart, even at opposite ends of the island: if you peel $YY, YB$ or $BY$, both bananas taste the same, with equal probability for 00 or 11, but if you peel $BB$, the bananas taste different, with equal probability for 01 or 10. I'll call these (E)PR pairs, because they are the analogue, in Bananaworld, of entangled  Einstein-Podolsky-Rosen (EPR) spin states or Bell states\footnote{The Bell states are  the four mutually orthogonal entangled pure states $\frac{1}{\sqrt{2}}(|0\rangle |1\rangle - |1\rangle |0\rangle), \frac{1}{\sqrt{2}}(|0\rangle |1\rangle +|1\rangle |0\rangle), \frac{1}{\sqrt{2}}(|0\rangle |0\rangle + |1\rangle |1\rangle), \frac{1}{\sqrt{2}}(|0\rangle |0\rangle - |1\rangle |1\rangle)$, related to each other  by local reversible operations. See \cite[p. 25]{NielsenChuang}.}   (the sort of quantum state considered by Einstein, Podolsky, and Rosen \cite{EPR} in their  argument for the incompleteness of quantum theory), but they exhibit the superquantum correlations of a PR-box. 

These correlations are impossible if each banana in an (E)PR pair has a  definite taste before it is peeled, assuming the taste of a banana is independent of how the paired banana is peeled (each of the four possible taste combinations 00, 01, 10, 11 is inconsistent with a possible pair of peelings). Moreover, it follows from Bell's locality argument \cite{BellEPR} that these correlations have no common cause explanation. 

A common cause (or local hidden variable representing `shared randomness') is a  variable $\lambda$  such that, given $\lambda$, the correlations vanish. That is, the joint probability of a pair of  tastes for a pair of peelings is conditionally statistically independent with respect to $\lambda$:
\begin{equation}
p(t_{1}, t_{2}|q_{1},q_{2},\lambda) = p(t_{1}|q_{1},\lambda)p(t_{2}|q_{2},\lambda) \label{eqn:cc}
\end{equation}
where $q_{1}$ and $q_{2}$ denote the type of peeling, $Y$ or $B$, for the two bananas, and $t_{1}$ and $t_{2}$ denote the respective tastes, ordinary (0) or intense (1), for the two bananas.

Conditional statistical independence is equivalent to the conjunction of two conditions:\footnote{The terminology is due to Abner Shimony \cite{Shimony1984}. For a proof of the equivalence here, see \cite[pp. 66--67]{Bubbook}.}
\begin {itemize}
\item \emph{parameter independence} (Bell locality, or no signaling given $\lambda$): the probability that an (E)PR banana tastes ordinary or intense, given $\lambda$ and a particular peeling ($Y$ or $B$) is independent of how you peel the paired banana
\item \emph{outcome independence}: the probability that an (E)PR banana tastes ordinary or intense, given $\lambda$ and a particular peeling ($Y$ or $B$), is independent of the taste of the paired banana after it is peeled
\end{itemize}

To see that the (E)PR correlations have no common cause explanation, it is convenient to re-label `ordinary' as -1 and `intense' as +1. Then, indicating the dependence on the common cause $\lambda$ by a subscript:
\begin{eqnarray}
\langle q_{1}q_{2}\rangle_{\lambda} & = &   p_{\lambda}(-1,-1|q_{1},q_{2}) - p_{\lambda}(-1,1|q_{1},q_{2}) - p_{\lambda}(1,-1|q_{1},q_{2}) \nonumber \\
& & \; \; + p_{\lambda}(1,1|q_{1},q_{2}) \label{eqn:joint} \\
& = &  p_{\lambda}(-1|q_{1})p_{\lambda}(-1|q_{2}) - p_{\lambda}(-1|q_{1})p_{\lambda}(1|q_{2}) - p_{\lambda}(1|q_{1})p_{\lambda}(-1|q_{2}) \nonumber \\
& & \; \;  + p_{\lambda}(1|q_{1})p_{\lambda}(1|q_{2}) \label{eqn:cc2} \\
& = &  (p_{\lambda}(1|q_{1}) - p_{\lambda}(-1|q_{1}))(p_{\lambda}(1|q_{2}) - p_{\lambda}(-1|q_{2})) \nonumber \\
& = &  \langle q_{1}\rangle_{\lambda} \langle q_{2}\rangle_{\lambda}
\end{eqnarray}
where the step from (\ref{eqn:joint}) to (\ref{eqn:cc2}) follows from conditional statistical independence (\ref{eqn:cc}).\footnote{The quantity $\langle q_{1}q_{2}\rangle$, the expectation value of the product of two peelings, like the expectation value of the product of two observables, is the weighted sum of the product of the tastes (represented numerically as -1, +1), where the weights are the joint probabilities of the pair of tastes. The product of the tastes is either 1 or -1, depending on whether the tastes are the same or different.}

So
\begin{eqnarray}
K_{\lambda} & = & \langle Y_{1}Y_{2}\rangle_{\lambda} +  \langle Y_{1}B_{2}\rangle_{\lambda} +  \langle B_{1}Y_{2}\rangle_{\lambda} -  \langle B_{1}B_{2}\rangle_{\lambda} \nonumber \\
& = & \langle Y_{1}\rangle_{\lambda} \langle Y_{2}\rangle_{\lambda} + \langle Y_{1}\rangle_{\lambda} \langle B_{2}\rangle_{\lambda} + \langle B_{1}\rangle_{\lambda} \langle Y_{2}\rangle_{\lambda} - \langle B_{1}\rangle_{\lambda} \langle B_{2}\rangle_{\lambda} \nonumber \\
& = &  \langle Y_{1}\rangle_{\lambda}(\langle Y_{2}\rangle_{\lambda} + \langle B_{2}\rangle_{\lambda}) + \langle B_{1}\rangle_{\lambda}(\langle Y_{2}\rangle_{\lambda} - \langle B_{2}\rangle_{\lambda})
\end{eqnarray}

Since the expectation values $\langle Y_{1}\rangle_{\lambda}, \langle Y_{2}\rangle_{\lambda}, \langle B_{1}\rangle_{\lambda}, \langle B_{2}\rangle_{\lambda}$ take the values $\pm 1$, the maximum and minimum values for either term in parenthesis is $+2$ or $-2$, in which case the other term in parenthesis is 0, so:
\begin{equation}
-2 \leq K_{\lambda} \leq 2
\end{equation}
It follow that $K = \langle YY \rangle + \langle YB\rangle + \langle BY\rangle - \langle BB\rangle = \int K_{\lambda}\rho(\lambda)d\lambda$, the Clauser-Horne-Shimony-Holt (CHSH) correlation (where $\rho(\lambda)$ is the distribution of common causes or local hidden variables), is similarly bounded:
\begin{equation}
-2 \leq K
 \leq 2
\end{equation}

But, since \begin{equation}
\langle q_{1}q_{2}\rangle = p(\mbox{taste same}|q_{1}, q_{2}) - p(\mbox{taste different}|q_{1},q_{2})
\end{equation} 
the CHSH correlation for (E)PR banana pairs is:
\begin{eqnarray}
K & = &  \langle YY\rangle + \langle YB\rangle + \langle BY\rangle - \langle BB\rangle \nonumber \\
& = &
p(\mbox{taste same}|Y,Y) + p(\mbox{taste same}|Y,B)  \nonumber \\
& & + p(\mbox{taste same}|B,Y) + p(\mbox{taste different}|B,B) \nonumber \\
& = & 4
\end{eqnarray}
which excludes a common cause or `shared randomness' explanation of (E)PR correlations. The maximum value of $K$ for a pair of entangled quantum states is $2\sqrt{2}$, the Tsirelson bound, which is between 2 and 4, so entangled quantum states are rather like (E)PR banana pairs.

A recent result by Pusey, Barrett, and Rudolph (PBR) \cite{PBR2012} shows that quantum pure states cannot be interpreted epistemically as states of knowledge represented by probability distributions over unknown `ontic' states. If quantum pure states could be interpreted epistemically, then any given ontic state $\lambda$ would be compatible with two or more different quantum pure states, in the sense that the distributions corresponding to the quantum states would overlap in a subset of the ontic state space containing $\lambda$. 

In the simple version of the argument, PBR consider preparing a quantum system (a qubit) in one of two quantum states: $|0\rangle$ or $|+\rangle = \frac{1}{\sqrt{2}}(|0\rangle + |1\rangle)$, where $|\langle 0|+\rangle| = \frac{1}{\sqrt{2}}$. Say $|0\rangle$ and $|1\rangle$ are eigenstates of $z$-spin, and $|+\rangle$ and $|-\rangle$ are eigenstates of $x$-spin. The supposition of the `no go' theorem is that the epistemic states $|0\rangle$ and $|+\rangle$ overlap in the  ontic state space.  There is therefore a certain finite probability $p$ that preparing the system in either of these states will result in an ontic state $\lambda$ in the overlap region. If, now, two independent non-interacting systems are prepared in either of the quantum states $|0\rangle$ or $|+\rangle$, with probability $p^{2}$ the systems will be in ontic states $\lambda_{1}, \lambda_{2}$, both in the overlap region. It follows that the ontic state $\lambda$ of the composite 2-qubit system (which the argument assumes is specified by the pair of local ontic states $\lambda = (\lambda_{1}, \lambda_{2})$, because the systems are independent and non-interacting) is compatible with the four quantum states: $|0\rangle\otimes |0\rangle, |0\rangle\otimes |+\rangle, |+\rangle\otimes |0\rangle,  |+\rangle\otimes |+\rangle$. 

Now PBR consider measuring an observable $A$  on the composite system with entangled eigenstates:
\begin{eqnarray}
\frac{1}{\sqrt{2}}(|0\rangle\otimes |1\rangle + |1\rangle\otimes |0\rangle) \\
\frac{1}{\sqrt{2}}(|0\rangle\otimes |-\rangle + |1\rangle\otimes |+\rangle) \\
\frac{1}{\sqrt{2}}(|+\rangle\otimes |1\rangle + |-\rangle\otimes |0\rangle) \\
\frac{1}{\sqrt{2}}(|+\rangle\otimes |-\rangle + |-\rangle\otimes |+\rangle)
\end{eqnarray}

The outcome of the measurement should depend only on the ontic state: given the ontic state, an epistemic state provides no further information relevant to the occurrence of an event. But, according to quantum mechanics, the outcome corresponding to $\frac{1}{\sqrt{2}}(|0\rangle\otimes |1\rangle + |1\rangle\otimes |0\rangle)$ is impossible (i.e., has zero probability) if the initial   quantum state is  $|0\rangle\otimes |0\rangle$, since $|0\rangle\otimes |0\rangle$ is orthogonal to $\frac{1}{\sqrt{2}}(|0\rangle\otimes |1\rangle + |1\rangle\otimes |0\rangle$. Similarly: 
\begin{itemize}
\item $|0\rangle\otimes |+\rangle$ is orthogonal to $\frac{1}{\sqrt{2}}(|0\rangle\otimes |-\rangle + |1\rangle\otimes |+\rangle)$
\item $|+\rangle\otimes |0\rangle$ is orthogonal to $\frac{1}{\sqrt{2}}(|+\rangle\otimes |1\rangle + |-\rangle\otimes |0\rangle) $
\item $|+\rangle\otimes |+\rangle$ is orthogonal to $\frac{1}{\sqrt{2}}(|+\rangle\otimes |-\rangle + |-\rangle\otimes |+\rangle)$
\end{itemize}
so none of the outcomes can occur. This is a contradiction.  The sum of the probabilities of the four possible outcomes is 1, so one of the outcomes must occur.

The conclusion PBR draw is that the quantum state must be real: `distinct quantum states must correspond to physically distinct states of reality' \cite[abstract]{PBR2011}, so a quantum state must be (at least part of) an ontic state  \cite[p. 476]{PBR2012}:
\begin{quote}
We have shown that the distributions for $|0\rangle$ and $|+\rangle$ cannot overlap. If the same can be shown for any pair of quantum states $|\psi_{0}\rangle$ and $|\psi_{1}\rangle$, then the quantum state can be inferred uniquely from $\lambda$. In this case, the quantum state is a physical property of the system.
\end{quote}

PBR proceed to prove this, from which it would seem that the only possible interpretative possibilities that take quantum mechanics as strictly true are Bohm's theory (in which the full ontic state is the quantum state together with the `hidden variable,' the position in configuration space of all the particles) or the Everett interpretation (in which the ontic state is just the quantum state). The general argument involves considering $n$ copies of the system, where $n$ depends on the angle between the states  $|\psi_{0}\rangle$ and $|\psi_{1}\rangle$ (the smaller the angle between the states, the larger the number $n$ of copies that needs to be considered). The argument does not require  the assumption of conditional statistical independence with respect to the ontic state for the quantum probabilities of joint events, so the result is different from Bell's theorem. What replaces conditional statistical independence is the (arguably innocuous) assumption that each of the copies can be prepared in a given pure quantum state independently of the others, and that this also applies to ontic states, so that there are no ÔsupercorrelationsÕ at the ontic level and the ontic state of the composite many-copy system is specified by the set of independent ontic states of the copies.

The PBR argument can be formulated in Bananaworld. First note that the four eigenstates of the observable $A$ that PBR measure in the simple version of the argument represent anticorrelation with respect to $z$-spin and $z$-spin, $z$-spin and $x$-spin, $x$-spin and $z$-spin, and $x$-spin and $x$-spin, respectively. With a bit of cheating (since no dynamics has been specified here for  Bananaworld), we will assume that the counterpart of the dynamical interaction associated with the $A$-measurement  involves putting the stems of two pure state bananas, each in one of the  states $Y0$ or $B0$, into a good quality espresso with small amount of foamed milk. We  find that they grow together---evolve dynamically---into a pair that is anticorrelated for one of the four possible peelings $YY, YB, BY, BB$, except that:
\begin{itemize}
\item if the pair is initially $Y0Y0$, then they never end up anticorrelated for $YY$
\item if the pair is initially $Y0B0$, then they never end up anticorrelated for $YB$
\item if the pair is initially $B0Y0$, then they never end up anticorrelated for $BY$
\item if the pair is initially $B0B0$, then they never end up anticorrelated for $BB$
\end{itemize}
This corresponds to the facts in the quantum case, e.g., the orthogonality of $|0\rangle\otimes |0\rangle$ and $\frac{1}{\sqrt{2}}(|0\rangle\otimes |1\rangle + |1\rangle\otimes |0\rangle)$, so that if the initial state is $|0\rangle\otimes |0\rangle$ then the outcome corresponding to $\frac{1}{\sqrt{2}}(|0\rangle\otimes |1\rangle + |1\rangle\otimes |0\rangle)$---anticorrelation with respect to the $z$-spins of the two particles---never occurs.

In a banana pure state like $Y0$, if you peel $B$, the banana tastes ordinary or intense with probability 1/2. So suppose $Y0$ is interpreted as an epistemic state, representing ignorance about some  ontic state $\lambda$. Then what happens if you put the stems of two pure state bananas into a macchiato depends on their ontic states, $\lambda_{1}, \lambda_{2}$. 

Suppose two epistemic states like $Y0, B0$ are compatible with the same ontic state. Then two bananas in the ontic states $\lambda_{1}, \lambda_{2}$, each compatible with the epistemic states $Y0, B0$, would be compatible with the four pairs of epistemic states: $Y0Y0, Y0B0, B0Y0, B0B0$.

But if that were the case, then contrary to what we find in Bananaworld, the two bananas could never become a pair that is anticorrelated for one of the four possible peelings $YY, YB, BY$, or $BB$, because the initial state $Y0Y0$ excludes eventual anticorrelation for $YY$, the initial state $Y0B0$ excludes eventual anticorrelation for $YB$, the initial state $B0Y0$ excludes eventual anticorrelation for $BY$, and the initial state $B0B0$ excludes eventual anticorrelation for $BB$.

A similar result by Colbeck and Renner \cite{ColbeckRenner2012}, from a very different argument, shows that no extension of quantum mechanics can improve predictability unless `free choice' is violated for measurements on a bipartite system in a pure entangled state: Alice's (or Bob's) choice of a measurement setting could not be independent of  information that is in principle available before the choice, i.e.,  the choice would be correlated with the ontic state prior to the choice in some frame of reference.  So, assuming `free choice,' it follows that quantum states are the whole story, i.e., quantum states are in 1-1 correspondence with ontic states. 

If `free choice' is violated for Alice, then Bob would have instantaneous information about Alice's choice of measurement setting if he has access to the ontic state immediately after Alice's measurement, so instantaneous signaling would be possible (unless access to the ontic state is blocked for some reason, as in Bohm's theory). The `no signaling' principle for quantum mechanics is the requirement that Alice's marginal probabilities, given the quantum state, are independent of Bob's measurement setting choices or, more generally, independent of anything Bob does, and conversely. Note that the no-signaling principle does not refer to any hidden variables or ontic states, and it does not involve the velocity of light, $c$. Of course if  the no-signaling principle is violated, then  the relativistic constraint is violated and superluminal signaling is possible.  But no signaling is a more fundamental principle than the relativistic constraint limiting the physical transfer of information to subluminal velocities---in effect, the satisfaction of the no-signaling principle for two systems is part of what it means to have two separate systems. Parameter independence is `no-signaling conditional on the ontic state'---or, equivalently, no-signaling  is parameter independence averaged over the ontic states. So a violation of `free choice' entails a violation of parameter independence, and conversely. 

In effect, the Colbeck-Renner result is a proof of Bell's theorem from parameter independence without requiring outcome independence, i.e., from a weaker assumption than Bell's assumption of  conditional statistical independence. Parameter independence is well-motivated because, assuming access to the ontic state, if this assumption were violated, then instantaneous signaling would be possible. Outcome independence does not have the same plausibility. Alice and Bob cannot exploit a violation of outcome independence to signal instantaneously, even given access to the ontic state, because measurement outcomes are random and not under their control: they can choose the measurement settings, but not the measurement outcomes. So it is an interesting question to consider whether one can prove a `no go' theorem from parameter independence without assuming outcome independence: is an extension of quantum mechanics with improved predictive power possible, where the extended theory satisfies parameter independence but not necessarily outcome independence?Ê Colbeck and Renner prove a result that answers the question negatively.
 
The Colbeck-Renner result has an analogue in Bananaworld. First note that the no-signaling principle is satisfied for (E)PR pairs: Alice's banana tastes ordinary or intense with equal probability, irrespective of  how Bob peels his banana, and conversely. Suppose  that Bob could somehow extract sufficient information from a measurement of some parameter $\lambda$ to know the taste of his banana for both possible peelings, $Y$ or $B$ (we can think of $\lambda$ as the ontic state of the (E)PR pair, or part of the ontic state). Suppose Bob's information is that the tastes for his two possible peelings are $j$ and $k$, where $j,k \in \{0,1\}$. Consider a reference frame in which Bob's measurement of $\lambda$ occurs immediately after Alice peels her banana.  If $j=k$, then Bob can infer that Alice peeled $Y$, because otherwise, if Alice peeled $B$, Alice's banana would have to taste the same as Bob's banana if he peeled $Y$, but different from Bob's banana if he peeled $B$, which is impossible if Bob's banana is in some ontic state that tastes the same for both possible peelings. Similarly, if $j \neq k$, then Bob can infer that Alice peeled $B$, because otherwise, if Alice peeled $Y$, Bob's banana for either of his two possible peelings would have to taste the same as Alice's banana, which would require Bob's banana to taste the same for either peeling. So Bob can infer whether Alice peeled $Y$ or $B$ from his measurement  of $\lambda$ and the no-signaling principle is violated: Alice can signal instantaneously to Bob. Alternatively, in a reference frame in which Bob's measurement of $\lambda$ occurs before Alice peels her banana,  Alice's measurement setting depends on  information that is available to Bob before her choice, so Alice's choice is not free. The same conclusion follows if the information Bob extracts from his measurement of $\lambda$ is partial: it suffices for him to discover whether the tastes for the two possible peelings of his banana are the same or different.

So there is a measurement problem in Bananaworld. If we reject the possibility of instantaneous signaling, or we assume that the choice of how to peel a banana is unconstrained or  free, in the sense that this choice is independent of any information that is in principle available before the choice, then there must be a \emph{necessary information loss} about whether an (E)PR banana peeled $Y$ would taste ordinary or intense in the course of a measurement that provides information about the taste if the banana is peeled $B$, and conversely. It follows that there can be no more refined story about the probabilistic correlations of an (E)PR pair  that yields information about the tastes of Alice's or Bob's banana for both possible peelings, or even information about whether the tastes are the same or different, unless `free choice' is violated or, equivalently, unless parameter independence (no signaling given the more refined story) is violated.

For similar reasons, it is impossible to clone a banana belonging to an (E)PR pair. Suppose Alice and Bob share an (E)PR pair of bananas, and Bob could clone his banana.  If Bob peels $Y$ for his banana and $B$ for the clone, then he could infer whether Alice peeled $Y$ or $B$ from the tastes of his bananas:  same taste indicates that Alice peeled $Y$; different taste indicates that Alice peeled $B$. So Alice could signal instantaneously to Bob if Bob peels his bananas immediately after Alice peels her banana (or superluminally if there is a suitably small time lag).  If Alice peels her banana  after Bob peeled his bananas, then Alice's choice of peeling depends on whether Bob's  banana and the clone taste the same or different. So if we assume that there is no constraint on the choice of how to peel a banana, or that there is no violation of the no-signaling principle, then cloning  an (E)PR banana is impossible.

The previous `no go' results---Bell, PBR, Colbeck-Renner---all involve correlations between spatially separate systems. But there are also correlations between the outcomes of compatible measurements on a single system. Kochen and Specker \cite{KochenSpecker} identified a finite  set of 1-dimensional projection operators on a 3-dimensional Hilbert space, in which an individual projection operator can belong to different orthogonal triples of projection operators representing different bases or measurement contexts, such that there is no noncontextual assignment of 0 and 1 values to the projection operators (where each projection operator is assigned one and only one value independent of context) that also respects the correlations defined by the orthogonality relations:  two orthogonal projection operators cannot both be assigned the value 1. It follows from the Kochen-Specker theorem that the distributions of measurement outcomes for quantum states in a Hilbert space of three or more dimensions cannot be simulated by noncontextual assignments of values to all observables, or even to certain finite sets of observables. 

The Kochen-Specker  proof is a state-independent proof. The simplest proof of the Kochen-Specker theorem is Klyachko's state-dependent proof \cite{Klyachko2002,Klyachko2007}.  Klyachko derived an inequality for probability assignments to the eigenvalues of five observables, which is satisfied  by any noncontextual assignment of values to this set of observables, but is violated by the probabilities defined by a certain quantum state. The argument is so pretty that it is worth repeating here.\footnote{This formulation of Klyachko's result, reproduced from  \cite{BubStairs2009},  owes much to a discussion with Ben Toner and differs from the analysis in \cite{Klyachko2002,Klyachko2007}.}

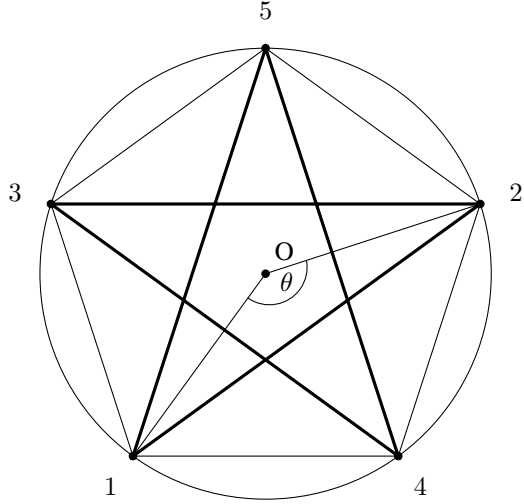
\begin{figure}[!ht]
    \begin{picture}(200,190)(-65,0)
\begin{tikzpicture}
\tikzstyle vertex=[circle,draw,fill=black,inner sep=1pt]
\path (0,0) coordinate (O);
\path (3*72+18:3cm) coordinate (P1);
\path (18:3cm) coordinate (P2);
\path (2*72+18:3cm) coordinate (P3);
\path (4*72+18:3cm) coordinate (P4);
\path (72+18:3cm) coordinate (P5);
\path (3*72+18:.4cm) coordinate (Q);
\node at (.25,.3)  {O};
\node at (-20:.3cm)  {$\theta$};
\path (3*72+18:3.5cm)
node {$1$};
\path (18:3.5cm) 
node {$2$};
\path (2*72+18:3.5cm) 
node {$3$};
\path (4*72+18:3.5cm)
node {$4$};
\path (72+18:3.5cm) 
node {$5$};
\draw[very thick] (P1) -- (P2) -- (P3) -- (P4) -- (P5) -- cycle;
\draw (P1) -- (P4) (P4) -- (P2)
(P2) -- (P5) (P5) -- (P3)
(P3) -- (P1);	
\draw (P1) -- (O) (O) -- (P2);
\draw (Q) arc (-125:10:.5cm);
\draw (0,0) circle (3cm);
\node[vertex] at (0,0) {};
\node[vertex] at (P1) {};
\node[vertex] at (P2) {};
\node[vertex] at (P3) {};
\node[vertex] at (P4) {};
\node[vertex] at (P5) {};
\end{tikzpicture}
    \label{fig:sigma1}
\end{picture}
\caption{Circle $\Sigma_{1}$}
 \end{figure}

\begin{figure}[!ht]
    \begin{picture}(300,250)(-80,0)
\begin{tikzpicture}
\tikzstyle vertex=[circle,draw,fill=black,inner sep=1pt]
\path (0,0) coordinate (P);
\path (0,-5) coordinate (O);
\path (3*72+18:3cm) coordinate (P1);
\path (18:3cm) coordinate (P2);
\path (2*72+18:3cm) coordinate (P3);
\path (4*72+18:3cm) coordinate (P4);
\path (72+18:3cm) coordinate (P5);
\path (0,-4.2) coordinate (Q);
\node at (.25,.3)  {P};
\node at (.3,-4.8)  {O};
\node at (-.2,-4.4)  {$\phi$};
\path (3*72+18:3.5cm)
node {$1$};
\path (3*72+10:1.2cm)
node {$s$};
\path (18:3.5cm) 
node {$2$};
\path (25:1.5cm) 
node {$s$};
\path (-94:2cm) 
node {$r$};
\path (-40:.65cm) 
node {$\sqrt{2}$};
\path (2*72+18:3.5cm) 
node {$3$};
\path (4*72+18:3.5cm)
node {$4$};
\path (72+18:3.5cm) 
node {$5$};
\draw[very thick] (P1) -- (P2) -- (P3) -- (P4) -- (P5) -- cycle;
\draw (P1) -- (P4) (P4) -- (P2)
(P2) -- (P5) (P5) -- (P3)
(P3) -- (P1);	
\draw (O) -- (P) (O) -- (P1) (P) -- (P2) (P) -- (P1); 
\draw (Q) arc (90:125:.8cm);
\draw (0,0) circle (3cm);
\node[vertex] at (0,0) {};
\node[vertex] at (P1) {};
\node[vertex] at (P2) {};
\node[vertex] at (P3) {};
\node[vertex] at (P4) {};
\node[vertex] at (P5) {};
\node[vertex] at (O) {};

\end{tikzpicture}

\end{picture}
\caption{Circle $\Sigma_{2}$}
    \label{fig:sigma2}
\end{figure}
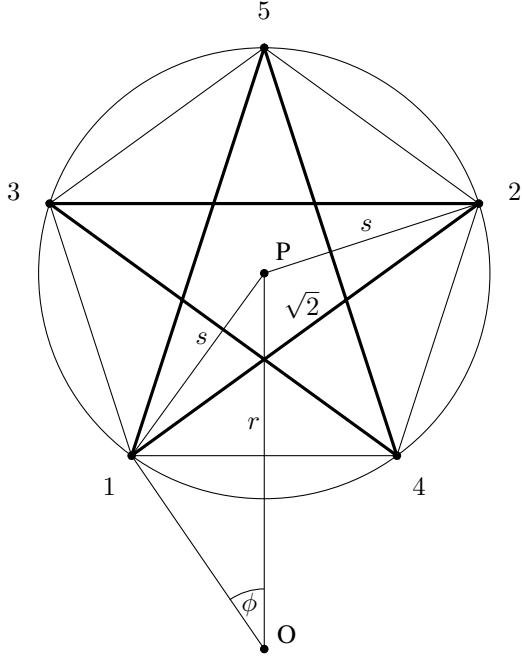

Consider a unit sphere and imagine a circle $\Sigma_{1}$ on the equator of the sphere with an inscribed pentagon and pentagram, with the vertices of the pentagram labelled in order 1, 2, 3, 4, 5 (see Figure \ref{fig:sigma1}). 
Note that the angle subtended at the center O by adjacent vertices of the pentagram defining an edge (e.g., 1 and 2) is $\theta = 4\pi/5$, which is greater than $\pi/2$. It follows that if the radii linking O to the vertices are pulled upwards towards the north pole of the sphere, the circle with the inscribed pentagon and pentagram will move up on the sphere towards the north pole. Since $\theta = 0$ when the radii point to the north pole (and the circle vanishes), $\theta$ must  pass through $\pi/2$ before the radii point to the north pole, which means that it is possible to draw a circle $\Sigma_{2}$ with an inscribed pentagon and pentagram on the sphere at some point between the equator and the north pole, \emph{such that the angle subtended at $O$ by an edge of the pentagram is $\pi/2$}.  Label the centre of \emph{this} circle $P$ (see Figure \ref{fig:sigma2}; note that the line OP is orthogonal to the circle $\Sigma_{2}$ and is not in the plane of the pentagram).

We can therefore define five orthogonal triples of vectors, i.e., five bases in a 3-dimensional Hilbert space $\hil{H}_{3}$, representing five different measurement contexts:
\[ \begin{array}{lll}
\ket{1},&\ket{2},&\ket{v} \\
\ket{2},& \ket{3}, & \ket{w}  \\
\ket{3}, & \ket{4}, & \ket{x} \\
\ket{4}, & \ket{5}, & \ket{y}  \\
\ket{5}, & \ket{1}, & \ket{z}
\end{array} \]
Here $\ket{v}$ is orthogonal to $\ket{1}$ and $\ket{2}$, etc. Note that each vector $\ket{1},\ket{2}, \ket{3}, \ket{4}, \ket{5}$ belongs to two different contexts. The vectors $\ket{u}, \ket{v}, \ket{x}, \ket{y}, \ket{z}$ play no role in the following analysis, and we can take a context as defined by an edge of the pentagram in the circle $\Sigma_{2}$.

Consider, now, assigning 0's and 1's to all the vertices of the pentagram in $\Sigma_{2}$ noncontextually (i.e., each vertex is assigned a value independently of the edge to which it belongs), in such a way as to satisfy the orthogonality constraint that at most one 1 can be assigned to the two vertices of an edge. The orthogonality constraint can be satisfied noncontextually by assignments of zero 1's, one 1, or two 1's (but not by three 1's, four 1's, or five 1's).  It follows that for any such noncontextual assignment, where $v(i)$ is the value assigned to the vertex $i$:
\begin{equation}
\sum_{i=1}^{5}v(i) \leq 2
\end{equation}
and so:
\begin{equation}
\sum_{i=1}^{5}p(v(i)=1) \leq 2
\end{equation}
for any distribution of noncontextual value assignments.

Now consider a quantum system in the state defined by a unit vector that passes through the north pole of the sphere. This vector passes through the point $P$ in the center of the circle $\Sigma_{2}$. Call this state $\ket{\psi}$. A simple geometric argument shows that if probabilities are assigned to the 1-dimensional projectors defined by the vertices of the pentagram on $\Sigma_{2}$ by the  state $\ket{\psi}$, then the sum of the probabilities is greater than 2.

To see this, note that the probability assigned to a vertex, say the vertex 1, is:
\begin{equation}
|\langle 1|\psi\rangle|^{2} = \cos^{2} \phi
\end{equation}
where $\ket{1}$ is the unit vector defined by the radius from O to the vertex 1. (Of course, each vertex is assigned the same probability by $|\psi\rangle$, which is symmetrically placed with respect to the five vertices, i.e., the angle $\phi$ is the same for each vertex.) Since the lines from the center $O$ of the sphere to the vertices of an edge of the pentagram on $\Sigma_{2}$ are radii of length 1 subtending a right angle, each edge of the pentagram has length $\sqrt{2}$. The angle subtended at $P$ by the lines joining $P$ to the two vertices of an edge is $4\pi/5$, so the length, $s$, of the line joining $P$ to a vertex of the pentagram is:
\begin{equation}
s = \frac{1}{\sqrt{2}\cos \frac{\pi}{10}}
\end{equation}
Now, $\cos \phi = r$, where $r$ is the length of the line $OP$, and $r^{2} + s^{2} = 1$, so:
\begin{equation}
\cos^{2} \phi = r^{2} = 1-s^{2} =  \frac{\cos \frac{\pi}{5}}{1+\cos\frac{\pi}{5}} = \frac{1}{\sqrt{5}}
\end{equation}
(because $\cos\pi/5 = \frac{1}{4}(1+\sqrt{5})$), and hence:
\begin{equation}
\sum_{i=1}^{5} p(v(i) =1) =  \sqrt{5} > 2
\end{equation}

There is an analogue of the Klyachko argument in Bananaworld. In addition to bananas with half the skin yellow and the other half brown, some banana trees have bunches of five bananas with perfectly ordinary yellow skins. The five bananas splay out from a central stem like the radii of a pentagram. If two adjacent bananas in these Klyachko bunches are peeled  while still on the stem, one banana always tastes ordinary (0) and the other banana intense (1).\footnote{We could have formulated the constraint for bananas connected by an edge of the pentagram, but it is simpler to consider the constraint for adjacent bananas, i.e., bananas connected by an edge of the pentagon. In effect, pentagon edges correspond to orthogonality relations for Klyachko bananas.} If the bananas are separated, peelings and tastes are uncorrelated. Once two adjacent bananas are peeled, the remaining three bananas turn out to be inedible. The bananas also turn out to be inedible if two non-adjacent bananas are peeled, or if more than two bananas are peeled.  

Labeling the bananas in order around the vertices of the pentagon (not the pentagram)  as $a,b,c,d,e$, it follows that:
\begin{eqnarray}
p(a\,\, \mbox{tastes}\,\,  1) + p(b\,\,  \mbox{tastes}\,\,  1) & = & 1 \nonumber \\
p(b\,\, \mbox{tastes}\,\,  1) + p(c\,\,  \mbox{tastes}\,\,  1) & = & 1 \nonumber \\
p(c\,\, \mbox{tastes}\,\,  1) + p(d\,\,  \mbox{tastes}\,\,  1) & = & 1 \nonumber \\
p(d\,\, \mbox{tastes}\,\,  1) + p(e\,\,  \mbox{tastes}\,\,  1) & = & 1 \nonumber \\
p(e\,\, \mbox{tastes}\,\,  1) + p(a\,\,  \mbox{tastes}\,\,  1) & = & 1 
\end{eqnarray}
So:
\[
2(p(a\,\, \mbox{tastes}\,\,  1) + p(b\,\, \mbox{tastes}\,\,  1) + p(c\,\, \mbox{tastes}\,\,  1) + p(d\,\, \mbox{tastes}\,\,  1) + p(e\,\, \mbox{tastes}\,\,  1))  =  5
\]
i.e.,
\begin{equation}
\sum_{i=a}^{e}p(i\,\, \mbox{tastes}\,\,  1) = \frac{5}{2} > 2
\end{equation}
in Bananaworld, which is the maximal violation of the Klyachko inequality consistent with the orthogonality constraint, just as the value 4 in Bananaworld is the maximal value of the CHSH inequality consistent with the no-signaling constraint. Labeling  the sum of the five probabilities as $\mathcal{K}$, we have:
\begin{equation}
\mathcal{K}_{\small\mbox{classical}} (= 2) < \mathcal{K}_{\small\mbox{quantum}} (= \sqrt{5} \approx 2.236) < \mathcal{K}_{\small\mbox{Bananaworld}} (= 2.5)
\end{equation}

\section{Taking Stock}

There is nothing intrinsically impossible about the phenomena in Bananaworld. A little weird, perhaps, but easy to imagine. What would we make of this? A banana split interpretation,  that each time one peels a banana, the banana and the entire Bananaworld with it, splits into multiple copies with differently peeled and different-tasting bananas, is not the first thing that leaps to mind.

Consider an (E)PR pair of bananas. The correlations can be represented in a correlation array as in Table \ref{Table:PR}. The probability $p(00|YY)$ is to be read as $p(0,0|Y,Y)$, i.e., as a joint conditional probability for both bananas tasting ordinary (0) when Alice and Bob both peel Y, and the probability $p(01|BY)$ is to be read as  $p(0,1|B,Y)$, etc. I drop the commas for ease of reading in this section; the first two slots in $p(--|--)$ before the conditionalization sign $|$ represent the two possible tastes for Alice's and Bob's bananas, respectively, and the second two slots after the conditionalization sign represent the two possible peelings that Alice and Bob choose, respectively. Note that sequences like 01 or $BY$ in this section represent pairs, not products. The marginal probability of 0 for Alice or  for Bob is obtained by adding the probabilities in the left column of each cell or the top row of each cell, respectively, and the marginal probability of 1  for Alice or  for Bob by adding the probabilities in the right column of each cell or the bottom row of each cell, respectively. The tastes are uncorrelated if the joint probability is expressible as a product of marginal or local probabilities for Alice and Bob; otherwise they are correlated. 
\begin{table}[h!]
\begin{center}
\begin{tabular}{|ll||ll|ll|} \hline
   &$\mbox{Alice}$&$Y$ & &$B$&\\
   $\mbox{Bob}$&&&&&\\\hline\hline
  $Y$ &&$p(00|YY) = \frac{1}{2}$&$ p(10|YY) = 0$  & $p(00|BY) = \frac{1}{2}$&$ p(10|BY) = 0$     \\
   &&$p(01|YY) = 0$&$p(11|YY) = \frac{1}{2}$  & $p(01|BY)=0$&$ p(11|BY) = \frac{1}{2}$  \\\hline
   $B$&&$p(00|YB)=\frac{1}{2}$&$ p(10|YB)=0$  & $p(00|BB)=0$&$ p(10|BB)=\frac{1}{2}$   \\
  &&$p(01|YB)=0$&$ p(11|YB)=\frac{1}{2}$  & $p(01|BB)=\frac{1}{2}$&$ p(11|BB)=0$   \\\hline
\end{tabular}
\end{center}
 \caption{(E)PR correlation array}
 \label{Table:PR}
  \end{table}

Now consider all possible correlation arrays of the above form. They  represent the pure and mixed states of a bipartite system with two binary-valued observables for each subsystem and form a regular convex polytope with 256 vertices, where the vertices represent the extremal deterministic arrays or pure states with probabilities $0$ or $1$ only.\footnote{A regular polytope is the multi-dimensional analogue of a regular polygon, e.g., an equilateral triangle, or a square, or a pentagon in two dimensions. A convex set is, roughly, a set such that from any point in the interior it is possible to `see' any point on the boundary.} The polytope is the closed convex hull of the vertices, i.e., the smallest closed convex set containing the vertices. There are four possible arrangements of $0$'s and $1$'s that add to $1$ in each square cell of the correlation array (i.e., one $1$ and three $0$'s), and four cells, hence $4^{4} = 256$ vertices. The 16 probability variables in the correlation array are constrained by the four probability constraints. It follows that the 256-vertex polytope is 12-dimensional.  A general correlation array  is represented by a point in this polytope, so the probabilities in the array can be expressed (in general, non-uniquely) as convex combinations of the $0, 1$ probabilities in extremal correlation arrays (just as the probability of one of two alternatives, $0$ or $1$, can be represented as a point on a line between the end points $0$ and $1$ because it can be expressed as a convex combination of the extremal end points).

There are 16 deterministic correlation arrays that satisfy the no-signaling principle, and the joint probabilities can all be expressed as products of marginal or local probabilities for Alice and Bob separately. For example, the deterministic correlation array in which the tastes are both ordinary (0) for all possible peeling combinations, as in Table \ref{Table:nonsignalingcorr}, is a no-signaling array and the joint probabilities can be expressed as a product of local probabilities: a marginal Alice-probability of 1 for the taste 0 given any peeling, and a marginal Bob-probability of 1 for the taste  0 given any peeling. 
 \begin{table}[h!]
\begin{center}
\begin{tabular}{|ll||ll|ll|} \hline
   &$\mbox{Alice}$&$Y$ & &$B$&\\
   $\mbox{Bob}$&&&&&\\\hline\hline
  $Y$ &&$p(00|YY) = 1$&$ p(10|YY) = 0$  & $p(00|BY) = 1$&$ p(10|BY) = 0$     \\
   &&$p(01|YY) = 0$&$p(11|YY) = 0$  & $p(01|BY)=0$&$ p(11|BY) = 0$  \\\hline
   $B$&&$p(00|YB)=1$& $p(10|YB)=0$  & $p(00|BB)=1$&$ p(10|BB)=0$   \\
  &&$p(01|YB)=0$&$ p(11|YB)=0$  & $p(01|BB)=0$&$ p(11|BB)=0$   \\\hline
\end{tabular}
\end{center}
 \caption{Extremal no-signaling deterministic correlation array}
 \label{Table:nonsignalingcorr}
\end{table}

The remaining 240 correlation arrays violate the no-signaling principle, and the joint probabilities cannot be expressed as products of marginal or local probabilities for Alice and Bob separately. For example, in Table \ref{Table:signalingcorr}, the taste of Alice's banana corresponds to Bob's choice of peeling: if Bob peels Y, Alice's banana tastes ordinary (0), and if Bob peels B, Alice's banana tastes intense (1). Similarly, the taste of Bob's banana corresponds to Alice's choice of peeling. So a choice of peeling by Alice or Bob is instantaneously revealed in  the taste of a remote banana. 
\begin{table}[h!]
\begin{center}
\begin{tabular}{|ll||ll|ll|} \hline
   &$\mbox{Alice}$&$Y$ & &$B$&\\
   $\mbox{Bob}$&&&&&\\\hline\hline
  $Y$ &&$p(00|YY) = 1$&$ p(10|YY) = 0$  & $p(00|BY) = 0$&$ p(10|BY) = 0$     \\
   &&$p(01|YY) = 0$&$p(11|YY) = 0$  & $p(01|BY)=1$&$ p(11|BY) = 0$  \\\hline
   $B$&&$p(00|YB)=0$& $p(10|YB)=1$  & $p(00|BB)=0$&$ p(10|BB)=0$   \\
  &&$p(01|YB)=0$&$ p(11|YB)=0$  & $p(01|BB)=0$&$ p(11|BB)=1$   \\\hline
\end{tabular}
\end{center}
 \caption{Extremal signaling deterministic correlation array}
 \label{Table:signalingcorr}
\end{table}

The 16 vertices defined by the local no-signaling deterministic correlation arrays are the vertices of a polytope: the polytope of local correlations. The local correlation polytope is included in a no-signaling nonlocal polytope, defined by the 16 vertices of the local polytope together with an additional 8 nonlocal vertices, one of these nonlocal vertices representing the standard (E)PR banana pair correlation array as defined above, and the other seven vertices representing  (E)PR banana pairs (also found in Bananaworld) related to  the standard (E)PR pair by relabeling Alice's peelings ($Y \rightarrow B, B \rightarrow Y$), and the tastes of Alice's banana ($0 \rightarrow 1, 1\rightarrow 0$) conditionally on Alice's choice of peeling, or Bob's peelings, and the tastes of Bob's banana   conditionally on  Bob's choice of peeling. For example, the correlations in Table \ref{Table:trprcorr} define an (E)PR pair.  Note that the 16 vertices of the local polytope can all be obtained from the vertex represented by Table \ref{Table:nonsignalingcorr} by similar local reversible operations. Both the 16-vertex local polytope and the 24-vertex no-signaling nonlocal polytope are 8-dimensional: in addition to the four probability constraints (one for each cell in the correlation array), there are four no-signaling constraints.
\begin{table}[h!]
\begin{center}
\begin{tabular}{|ll||ll|ll|} \hline
   &$\mbox{Alice}$&$Y$ & &$B$&\\
   $\mbox{Bob}$&&&&&\\\hline\hline
  $Y$ &&$p(00|YY) = 0$&$ p(10|YY) = \frac{1}{2}$  & $p(00|BY) = 0$&$ p(10|BY) = \frac{1}{2}$     \\
   &&$p(01|YY) =\frac{1}{2}$&$p(11|YY) = 0$  & $p(01|BY)=\frac{1}{2}$&$ p(11|BY) = 0$  \\\hline
   $B$&&$p(00|YB)=0$& $p(10|YB)=\frac{1}{2}$  & $p(00|BB)=\frac{1}{2}$&$ p(10|BB)=0$   \\
  &&$p(01|YB)=\frac{1}{2}$&$ p(11|YB)=0$  & $p(01|BB)=0$&$ p(11|BB)=\frac{1}{2}$   \\\hline
\end{tabular}
\end{center}
 \caption{(E)PR correlation array related to the standard (E)PR correlation array in Table~\ref{Table:PR} by relabeling}
\label{Table:trprcorr}
\end{table}

Correlations represented by points in the local polytope can be simulated by Alice and Bob with classical resources, which generate classical correlations represented by the points in a simplex, where the vertices represent extremal states that are the common causes of the correlations: joint `local hidden variable' deterministic states expressing shared randomness.  A simplex is a regular convex polytope generated by $n+1$ vertices that are not confined to any $(n-1)$-dimensional subspace, e.g., a tetrahedron as opposed to a square. The lattice of subspaces of a simplex (the lattice of vertices, edges, and faces) is a Boolean algebra, with a 1-1 correspondence between the vertices, corresponding to the atoms of the Boolean algebra, and the facets (the $(n-1)$-dimensional faces), which correspond to the co-atoms. The classical simplex---in this case a 16-vertex  simplex---represents the correlation polytope of probabilistic states of two Bananaworld bananas that  behave `classically,' i.e., like a bipartite classical system with two binary-valued observables for each subsystem; the associated Boolean algebra represents the classical possibility structure. Probability distributions over these extremal states---mixed states---are represented by points in the interior or on the boundary of the simplex.

\begin{figure}[!h]
\begin{picture}(200,140)(20,40)
\scalebox{1.4}{\includegraphics{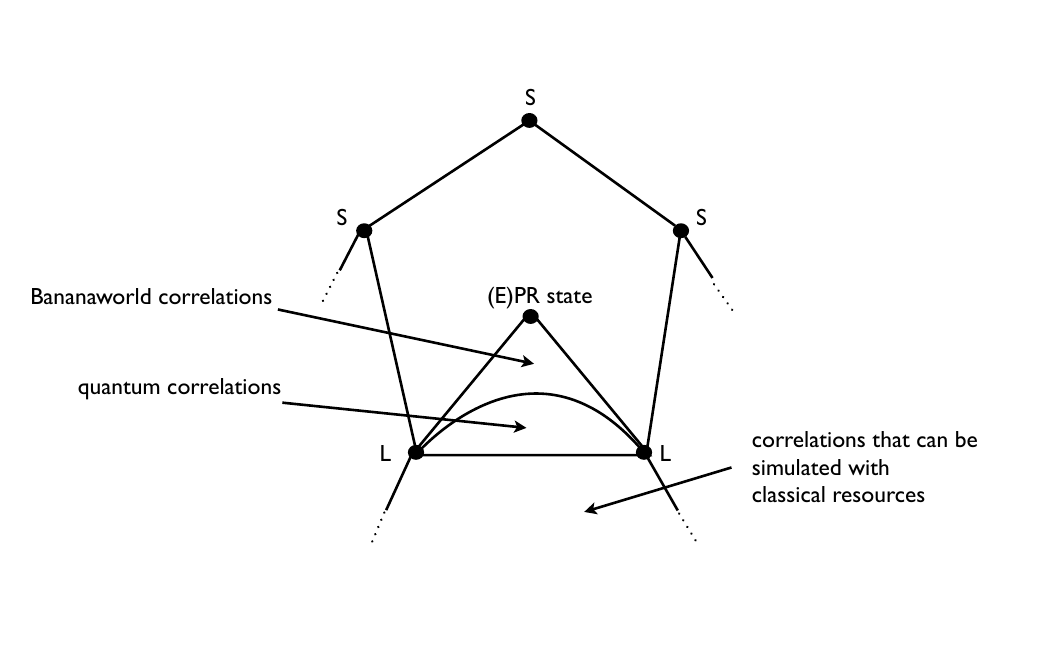}}
\end{picture}
\caption[]{Schematic representation of part of the space of all possible correlation arrays for two bananas with two possible peelings, $Y$ or $B$, and two possible tastes, ordinary (0) or intense (1). The vertices L are the deterministic vertices of the local no-signaling polytope. Correlations represented by points in this polytope can be simulated with classical resources.  The vertices S are the nonlocal signaling vertices (240 vertices). The Bananaworld polytope is bounded by the sixteen vertices L of the local no-signaling polytope together with eight (E)PR vertices, which represent the strongest possible nonlocal correlations consistent with the no-signaling principle. The hyperspherical quantum convex set is bounded by a continuous set of vertices and is not a polytope. It lies between the 16-vertex local polytope and the 24-vertex Bananaworld no-signaling polytope.} 
\label{Fig:all}
\end{figure}

The 24-vertex no-signaling nonlocal polytope is included in the 256-vertex nonlocal polytope with 240 vertices that represent deterministic signaling states. The correlations of a bipartite quantum system with  two binary-valued observables  for each subsystem are bounded by the Tsirelson bound and are represented by a hyperspherical convex set that  is not a polytope, with a continuous set of extremal points (quantum  pure states) between the 16-vertex local polytope and the 24-vertex no-signaling nonlocal polytope. See Figure \ref{Fig:all}. Note that the vertices L of the local correlation polytope are included in the quantum convex set---a correlation represented by L can be simulated (trivially) by Alice and Bob with shared quantum states.

A simplex has the rather special property that a mixed state, represented by a point in the interior of the simplex, can be expressed \emph{uniquely} as a mixture (convex combination) of extremal or pure states, the vertices of the simplex. \emph{No other polytope or convex set has this feature.} So in the class of no-signaling theories, classical (= simplex) theories are rather special. If there is no unique decomposition of mixed states into deterministic pure states for the operational or phenomenal probabilities, as is the case for the local 16-vertex polytope, then either  there is something left out of the story (which could, in principle, be added to complete the story as a simplex theory in which the vertices represent the common causes of correlations), or the correlations are outside the local polytope and there is no theoretical account that provides an explanation of correlations in terms of deterministic pure states without violating the no-signaling principle: any no-signaling explanation of correlations will have to include indeterministic states like (E)PR states. 

Fundamentally, the conceptual problems in Bananaworld, as in our quantum world, arise because probabilistic correlations  can lie outside the local correlation polytope. The existence of correlations outside the local correlation polytope means that we have to give up the view that a  banana in Bananaworld has what Einstein called a `being-thus' (`So-Sein' in German) \cite[p. 170]{Einstein1948}, in the sense that a banana is characterized by definite properties, prior to any peeling, that are correlated, either deterministically or probabilistically, with a particular taste for a particular peeling.
\begin{quotation}
\noindent If one asks what, irrespective of quantum mechanics, is characteristic of the world of ideas of physics, one is first of all struck by the following: the concepts of physics relate to a real outside world, that is, ideas are established relating to things such as bodies, fields, etc., which claim a `real existence' that is independent of the perceiving subject---ideas which, on the other hand, have been brought into as secure a relationship as possible with sense-data. It is further characteristic of these physical objects that they are thought of as arranged in a physical space-time continuum.  An essential aspect of this arrangement of things in physics is that they lay claim, at a certain time, to an existence independent of one another, provided that these objects `are situated in different parts of space.' Unless one makes this kind of assumption about the independence of the existence (the `being-thus') of objects which are far apart from one another in space---which stems in the first place from everyday thinking---physical thinking in the familiar sense would not be possible. It is also hard to see any way of formulating and testing the laws of physics unless one makes a clear distinction of this kind.\footnote{The English translation is due to Max Born, and the reference here is to the translation in \cite{Born}.}
\end{quotation}

To be sure, `physical thinking in the familiar sense' will have to change, but formulating and testing the laws of Bananaworld does not present any special problem.  The bananas in Bananaworld can be peeled and tasted. The phenomena are objective. We don't know how the universe makes  bananas with Bananaworld correlations, but we don't know how the universe makes objects with independent `being-thuses' either. What we know is how elementary objects with elementary `being-thuses' can be composed or transformed dynamically into complex objects with complex `being-thuses,' and in this sense we understand how correlations arise among such objects. The situation in Bananaworld (or our quantum world) is no different: given elementary entities such as pure state bananas or (E)PR pairs, we can understand the behavior of complex systems. 

\section{The Measurement Problem}

What of the measurement problem in Bananaworld? The appropriate answer to the question about how a banana selects a particular taste for a particular peeling---how `the transition from the possible to the actual' comes about---is that this occurs freely, as a truly random event, consistent with the probabilistic correlations of the Bananaworld correlation polytope. There is nothing more that needs to be said or can be said about how the trick is done.  In a simplex theory, it is possible to give a dynamical account of the transition from one actually occurring event to another event,  or to explain how a particular measurement outcome is selected as a dynamical transition from one vertex of the simplex to another. In a non-simplex theory, there is no analogous dynamical account, because any such account would be inconsistent with `free choice' or the  no-signaling principle---it would have to involve signaling states. 

There is, however, a remaining measurement problem in Bananaworld. Here is how Bohm puts the  problem for quantum mechanics \cite[p. 583]{Bohm}:
\begin{quotation}
\noindent If the quantum theory is to be able to provide a complete description of everything that can happen in the world \ldots it should also be able to describe the process of observation itself in terms of the wave functions of the observing apparatus and those of the system under observation. Furthermore, in principle, it ought to be able to describe the human investigator as he looks at the observing apparatus and learns what the results of the experiment are, this time in terms of the wave functions of the various atoms that make up the investigator, as well as those of the observing apparatus and the system under observation. In other words, the quantum theory could not be regarded as a complete logical system unless it contained within it a prescription in principle for how all these problems were to be dealt with. 
\end{quotation}

The bananas in Bananaworld  are wildly nonclassical (and nonquantum) objects, but the presumption is that visitors to the island are classical beings who can make \emph{definite} choices about how to peel bananas, and who have \emph{definite} taste sensations when they eat a peeled banana. The banana trees are located in \emph{definite} locations on the island, and have \emph{definite} bunches of bananas of certain sorts, and so on. If we want a unified physics, we can't simply assume that all these systems have definite `being-thuses,' while only the bananas exhibit correlations outside the local correlation polytope. What if, in effect, `everything is made of bananas,' i.e., if everything is made of systems with correlations that can lie outside the local correlation polytope, so that no system, including visiting observers to Bananaworld, can be understood as having a definite `being-thus' in Einstein's sense? Unless there is a response to Bohm's question, any theory of Bananaworld phenomena can only be an instrumentalist theory, where  observers and their measuring instruments and recording devices are left out of the theoretical story. What's wrong with the Copenhagen interpretation is precisely the assumption that observers are like visitors to Bananaworld, and that quantum phenomena, like banana phenomena, take place in a classical arena that provides the physical infrastructure for definite stable records. Bohr's defense of this position \cite[p. 209]{Bohr1949}:
\begin{quotation}
\noindent The argument is simply that by the word \emph{experiment} we refer to a situation where we can tell others what we have done and what we have learned and that, therefore, the account of the experimental arrangements and the results of the observations must be expressed in unambiguous language with suitable application of the terminology of classical physics,
\end{quotation}
is hardly more than  an excuse for leaving observers and the recording infrastructure out of the quantum theoretical story. 

Can we show that a dynamics for Bananaworld is \emph{consistent} with the assumption that observers and their instruments (fingers, tastebuds, etc.) always have definite properties, so that  we are entitled to infer from our theoretical account that something definite or determinate happens when  a banana is peeled and tasted, or in general when we make a measurement? In the case of Bananaworld, this is doubtful, because of constraints on a reasonable dynamics (see, e.g., \cite{Barrett2007}). 

For quantum mechanics, what we have to show is that the dynamics, which generally produces entanglement between two coupled systems, is consistent with the assumption that something definite happens in a measurement process. The basic question is whether it is consistent with the unitary dynamics to take the macroscopic measurement `pointer' or, in general, the macroworld as definite. The answer is `no,' if we accept an  interpretative principle sometimes referred to as the `eigenvalue-eigenstate link.' 

Dirac's states  the principle as follows \cite[pp. 46--47]{Dirac}\footnote{ One finds similar remarks in von Neumann \cite[p. 253]{Neumann} and in the EPR paper \cite{EPR}. The EPR argument is formulated as a \emph{reductio} for the principle: EPR show that it follows from the principle, together with certain realist assumptions, that quantum mechanics is incomplete}:
\begin{quote}
\noindent The expression that an observable `has a particular value' for a particular state is permissible in quantum mechanics in the special case when a measurement of the observable is certain to lead to the particular value, so that the state is an eigenstate of the observable. \ldots In the general case we cannot speak of an observable having a value for a particular state, but we can speak of its having an average value for the state. We can go further and speak of the probability of its having any specified value for the state, meaning the probability of this specified value being obtained when one makes a measurement of the observable.
\end{quote}

Dirac's principle, together with the linearity of the unitary dynamics of quantum mechanics, leads to the measurement problem: the definiteness of a particular measurement outcome is inconsistent with the entangled state that linearity requires at the end of a measurement. Typically, the measuring instrument is a macroscopic system, like Schr\"{o}dinger's cat, where the cat states $\ket{\mbox{alive}}$ and $\ket{\mbox{dead}}$ are correlated with eigenstates of a microsystem in the final entangled state at the end of the dynamical interaction. Here the macroscopic cat states act as measurement pointer states for the eigenvalues of the measured observable of the microsystem.

The eigenvalue-eigenstate link is an interpretative principle, a \emph{stipulation} about when an observable `has a particular value,' that is not  required by the kinematic structure of quantum mechanics. Alternative stipulations are possible. In particular, no contradiction is involved in stipulating that some particular `preferred' observable, $R$, is always definite (as long as we don't stipulate other noncommuting observables as also definite). 

The following theorem \cite{BubClifton} characterizes all possible stipulations, subject to some minimal constraints. If $R$ is a preferred observable in this sense (stipulated as always definite, i.e., as always having a determinate value), and $e$ is the ray representing a pure quantum state, there is a unique maximal sublattice $\mathcal{L}(R,e)$ in the lattice of Hilbert space subspaces representing quantum propositions, such that:
\begin{itemize}
\item  $\mathcal{L}(R,e)$ contains the eigenspaces of the preferred observable $R$
\item  $\mathcal{L}(R,e)$ is defined by $R$ and $e$ alone (i.e., $L(R,e)$ is invariant under Hilbert space lattice automorphisms that preserve $R$ and $e$)
\item $\mathcal{L}(R,e)$ is an ortholattice admitting sufficiently many 2-valued homomorphisms (truth-value assignments)\footnote{But not sufficiently many to distinguish every pair of distinct elements, as there is in a Boolean algebra.}  to recover all the single and joint probabilities defined by the state $e$ for mutually compatible sets of elements in $\mathcal{L}(R,e)$ as measures on a Kolmogorov probability space, so that the probability of an element $a$ in  $\mathcal{L}(R,e)$ is just the measure of the set of 2-valued homomorphisms that assign 1 to $a$
\end{itemize}

See \cite{HalvorsonClifton1999} for a $C^{*}$-algebraic generalization of the theorem, and \cite{Nakayama2008} for a topos-theoretic generalization. The elements in $\mathcal{L}(R,e)$ are obtained by projecting $e$ onto the eigenspaces of $R$. These projections, together with all the rays in the subspace orthogonal to the span of the projections, generate $\mathcal{L}(R,e)$. In the case of Schr\"{o}dinger's cat, if we take $R$ as the cat-observable with eigenstates $|\mbox{alive}\rangle$ and $|\mbox{dead}\rangle$, then it follows that the cat is definitely alive or definitely dead in the entangled state after the interaction with the microsystem, and the microsystem property correlated with the cat being definitely alive or the cat being definitely dead is also definite in the final entangled state.

Here is one way to solve the consistency problem: If we take $R$ as the decoherence `pointer' selected by environmental decoherence, then it follows that the macroworld is always definite because of  the nature of the decoherence interaction coupling environmental degrees of freedom to macroworld degrees of freedom (a contingent feature of the quantum dynamics), and it follows from the theorem that features of the microworld correlated with $R$ are definite. In other words, decoherence guarantees the continued definiteness or persistent objectivity of the macroworld, if we stipulate that $R$ is the decoherence `pointer.'

The eigenvalue-eigenstate link, or Dirac's principle, amounts to the stipulation that the preferred observable $R$ is the identity $I$, which has the whole Hilbert space and the null space as the two eigenspaces. There is a sense in which the eigenstate-eigenvalue link is preserved by the alternative  stipulation of $R$ as the decoherence pointer. If a system $S$ is not entangled with the decoherence pointer $R$, so that the total state is a product state $|s\rangle\otimes \cdots$, where $|s\rangle$ is the state of $S$, then the decoherence pointer takes the form $I\otimes R$, where $I$ is the identity in the Hilbert space of $S$, and the definite $S$-properties according to the above theorem are as specified by the eigenstate-eigenvalue link.

The argument here is not that decoherence provides a dynamical explanation of how an indefinite quantity becomes definite in a measurement process---Bell \cite{BellCH} has aptly criticized this argument as a `for all practical purposes' (FAPP) solution to the measurement problem. Rather, the claim is that we can take the decoherence pointer as definite \emph{by stipulation}, and that decoherence then guarantees the objectivity of the macroworld, which resolves the measurement problem without resorting to Copenhagen or neo-Copenhagen instrumentalism.

To sum up: What we have discovered is that there are probabilistic correlations outside the polytope of local correlations---the structure of information is not what we thought it was. The `no go' theorems tell us that we can't shoe-horn these correlations into a classical simplex by supposing that something has been left out of the story, without giving up fundamental principles that define what we mean by a physical system. The nonclassical features of quantum mechanics, including the irreducible information loss on measurement, are generic features of non-simplex theories. 

As far as the conceptual problems are concerned, we might as well talk about Bananaworld, bearing in mind that what counts as a good explanation in a non-simplex theory should not be constrained by the standard ontology that goes along with explanation in a simplex theory. In particular, we should not expect to find a dynamical explanation for the selection of a particular outcome in a quantum measurement, which is a genuinely random event, a `free choice' on the part of a quantum system. The complementary information loss or `irreducible and uncontrollable measurement disturbance' is a kinematic feature of  \emph{any} process
of gaining information of the relevant sort, irrespective of the particular unitary dynamical process involved in the
measurement process---just as the phenomenon of  Lorentz contraction is 
explained relativistically as a kinematic effect of motion in a
non-Newtonian space-time structure, irrespective of the particular Lorentz covariant dynamical process associated with the specific material constitution of the contracting system. 

The kinematic part of quantum mechanics as a non-simplex theory is given by the subspace structure of Hilbert space, which represents the structure of the space of possibilities for quantum events. This includes the association of  Hermitian
operators with observables, the Born probabilities, the von Neumann-L\"{u}ders
conditionalization rule, and the unitarity constraint on the dynamics, which is related to the possibility structure via a theorem of Wigner \cite{Wigner1959,Uhlhorn1963}. The possibility space is a non-Boolean space in which there are built-in, structural probabilistic constraints on correlations between events (associated with the angles between the rays representing extremal events)---just as in special
relativity the geometry of Minkowski space-time represents spatio-temporal
constraints on events. These are kinematic, i.e., pre-dynamic, objective probabilistic or information-theoretic constraints
on events to which a quantum dynamics of matter and fields conforms, through its symmetries, just
as the structure of  Minkowski space-time imposes spatio-temporal kinematic constraints on events to which a relativistic dynamics conforms.  

The  problem of how to account for the definiteness or determinateness of the part of the universe that records the outcomes of quantum measurements or random quantum events is a consistency problem. The question is whether  it is consistent with the quantum dynamics to take some part of the universe, including the registration of quantum events by our macroscopic measuring instruments, as having a definite `being-thus,' characterized by definite properties. This is not so much a conceptual problem as a problem for physics. The phenomenon of decoherence, a contingent feature of the system-environment interaction in our quantum universe, provides a solution.

\section*{Acknowledgements}
My research is supported by the Institute for Physical Science and Technology at the University of Maryland. This publication was made possible through the support of a grant from the John Templeton Foundation. The opinions expressed in this publication are those of the author and do not necessarily reflect the views of the John Templeton Foundation. Discussions with Allen Stairs and James Mattingly are gratefully acknowledged. Special thanks to Tony Sudbery for clarification about polytopes. 

\bibliographystyle{plain}
\bibliography{bworld.bib}

\end{document}